\documentclass[%
 sor,
 jor,
 amsmath,amssymb,
 reprint,%
author-numerical,%
]{revtex4-2}

\usepackage{graphicx}
\usepackage{dcolumn}
\usepackage{bm}
\usepackage{pifont}

\begin{document}

\preprint{APS/123-QED}

\title{Physics-driven generative adversarial networks empower single-pixel infrared hyperspectral imaging}

\author{Dong-Yin Wang}
\author{Shu-Hang Bie}%

\author{Xi-Hao Chen}
\email{xi-haochen@163.com}
\affiliation{%
 Key Laboratory of Optoelectronic Devices and Detection Technology, School of Physics, Liaoning University, Shenyang 110036, China
}%


\author{Wen-Kai Yu}
\email{yuwenkai@bit.edu.cn}
\affiliation{
 Center for Quantum Technology Research, and Key Laboratory of Advanced Optoelectronic Quantum Architecture and Measurement of Ministry of
Education, School of Physics, Beijing Institute of Technology, Beijing 100081, China}%

\date{\today}

\begin{abstract}
A physics-driven generative adversarial network (GAN) was established here for single-pixel hyperspectral imaging (HSI) in the infrared spectrum, to eliminate the extensive data training work required by traditional data-driven model. Within the GAN framework, the physical process of single-pixel imaging (SPI) was integrated into the generator, and the actual and estimated one-dimensional (1D) bucket signals were employed as constraints in the objective function to update the network's parameters and optimize the generator with the assistance of the discriminator. In comparison to single-pixel infrared HSI methods based on compressed sensing and physics-driven convolution neural networks, our physics-driven GAN-based single-pixel infrared HSI can achieve higher imaging performance but with fewer measurements. We believe that this physics-driven GAN will promote practical applications of computational imaging, especially various SPI-based techniques.
\end{abstract}

\maketitle

\section{\label{sec:level1}Introduction}
Deep learning (DL) \cite{1sarker2021deep}, originated from machine learning \cite{2jordan2015machine}, aims to mimic human neural networks for intelligent tasks. Nowadays, after several decades' development, it finds applications in healthcare \cite{3miotto2018deep,4rahman2023federated}, natural language processing \cite{5lauriola2022introduction}, transportation \cite{6kashyap2022traffic}, science and technology \cite{7choudhary2022recent}, and the like. Recently, the integration of DL with various optical imaging techniques has opened up extensive prospects for intelligent perception, significantly enhancing the imaging performance of both traditional and newly emerging imaging methods, including micrographic imaging \cite{xianweiragone2023deep}, ghost imaging (GI) \cite{8lyu2017deep}, and single-pixel imaging (SPI) \cite{9song2023single,33lyu2017deep,34hoshi2022single,35liu2023low,36rizvi2020deep,37rizvi2019improving}. In the field of optical imaging, the primary DL networks can be classified into two categories based on their drive modes: data-driven (trained) and physics-driven (untrained) neural networks. The former were initially introduced into various imaging and image processing fields, showing excellent performance, especially in GI or SPI \cite{33lyu2017deep,34hoshi2022single}. However, these data-driven networks require a large amount of input and output data pairs for training, which results in inherent issues such as generalization \cite{fanhuaalzubaidi2023survey,fanhuabrigato2020close}, interpretability, and very long model training time. To address these issues existing in data-driven networks, the physics-driven networks, which are derived from the former and the theory of deep image priors (DIP) \cite{51ulyanov2018deep}, have been introduced. The theory of DIP states that a meticulously designed neural network with randomly initialized weights possesses a prior capability biased toward natural images. Currently, this approach has been extensively applied to a variety of optical imaging scenarios \cite{52wang2020phase,53lin2022compressed,54gelvez2023mixture,55bostan2020deep}. It not only resolves the problem of catastrophic forgetting, but also further improves interpretability. Additionally, it can also alleviates the issue of generalization in optical imaging at a low computational cost, and save the data training time, showcasing promising potentials.

Hyperspectral imaging (HSI) is an advanced imaging technique that typically employs multiple well-defined optical bands in a wide spectral range to capture object images, thereby acquiring a set of two-dimensional (2D) images at different wavelengths \cite{yuanligarini2006spectral}, providing a richer and more extensive spectral information compared to red-green-blue (RGB) and multispectral imaging. Over the last two decades, HSI has transitioned from its initial use in remote sensing via satellites and aerial platforms to a wide range of applications, including encompassing mineral exploration \cite{14deng2021identification}, medical diagnostics \cite{yixuebarberio2021intraoperative}, environmental monitoring \cite{15stuart2019hyperspectral}, and the like. However, the need for high-resolution images in HSI to capture fine scene details inevitably leads to substantial data acquisition, increased sampling time, and higher processing and storage costs. The introduction of compressed sensing (CS) algorithms and SPI provides one of the promising alternative solutions for HSI \cite{wang2023single,25hmagalhaes2012high,26hwelsh2013fast,27radwell2014single,28august2013compressive,29hahn2014compressive,30tao2021compressive,31hyi2020hadamard}, effectively addressing the challenges associated with HSI by utilizing subsampling and single-pixel detection. An illustrative study has demonstrated the effectiveness of a CS-based single-pixel HSI method for detecting the chemical composition of targets in the near-infrared spectrum. It highlights the method's potential for efficient and cost-effective chemical composition detection through high-compression spectral analysis \cite{huaxuegattinger2019broadband}. The other approach is to introduce deep neural networks on the foundation of CS-based SPI and other techniques \cite{wang2023single,dlcmur2020deep,46hkim2017hazardous,47hheiser2019compressive,48hitasaka2019dnn,49hxie2019hyperspectral,491li2011compressive,492magalhaes2012high,493welsh2013fast,494hahn2014compressive,495tao2021compressive,496yi2020hadamard}. Inspired by the first GI scheme using a physics-driven deep CNN constraint (e.g., GIDC) \cite{50GIDCwang2022}, a physics-driven convolutional neural network (CNN) was integrated with a single-pixel HSI setup to achieve high-quality reconstruction across a wide range of visible wavelengths at a low sampling rate \cite{wang2023single}. However, CNNs have limitations that their accuracy of the learning is not high enough. Therefore, there is an urgent need to investigate physics-driven state-of-the-art networks for further improving the performance of HSI.

In this paper, a physics-driven approach based on generative adversarial network (GAN) was proposed for single-pixel infrared HSI by effectively integrating the physical process of SPI and network units of GAN, resulting in the elimination of extensive dataset training required in traditional data-driven deep single-pixel HSI. In this work, a comparison among our method, the CS algorithm and physics-driven CNN (i.e., GIDC) was made through both numerical simulations and optical experiments, revealing that better imaging quality of the recovered infrared hyperspetral images can be achieved at lower sampling rates (SR) by using our approach.

\section{\label{sec:level1}Principle and method}
\subsection{\label{sec:level2}Experimental setup}
\begin{figure}[htbp]
\centering
\includegraphics[width=1\linewidth]{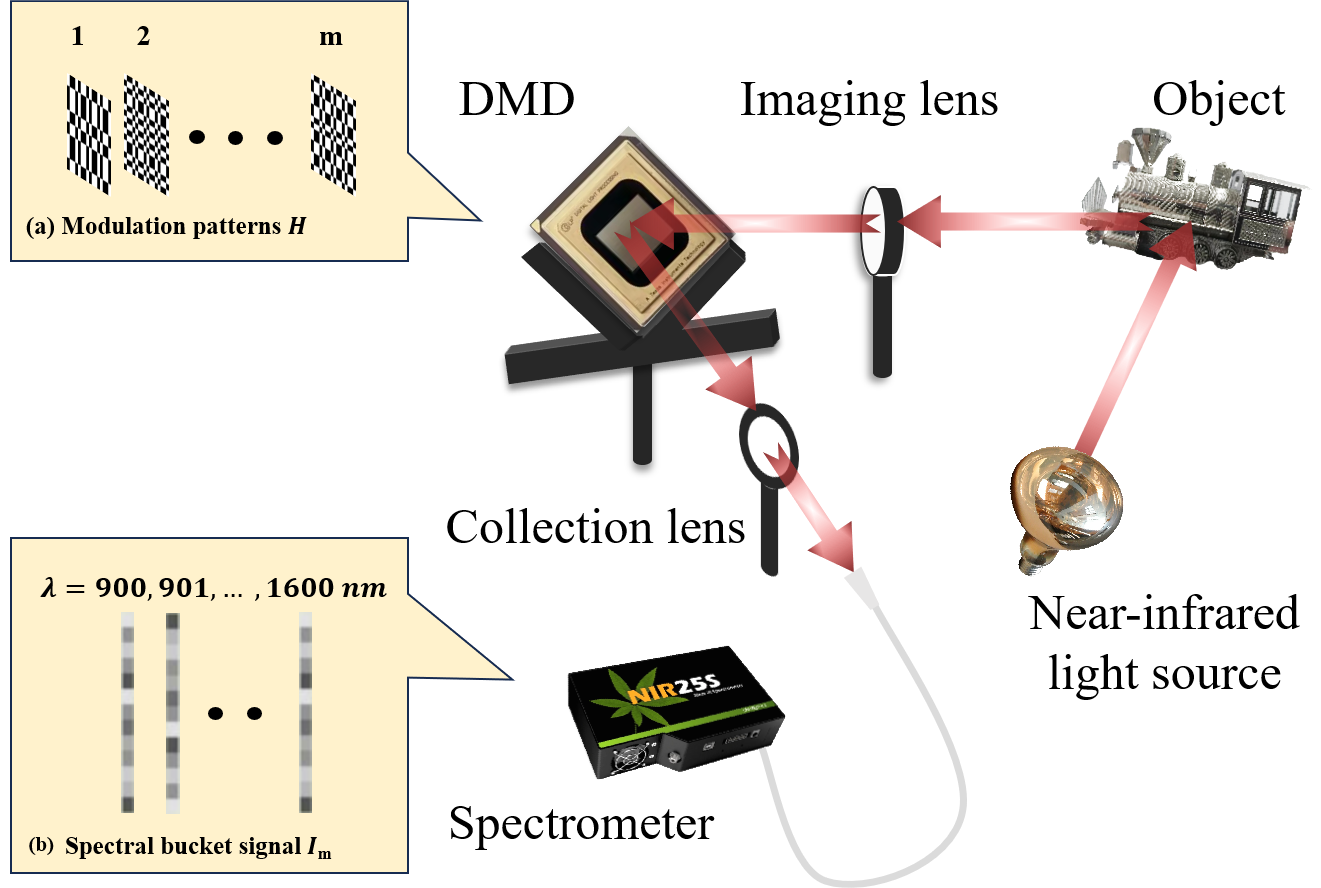}
\caption{Schematic diagram of the experimental setup. (a) Modulation pattern $H_m$. (b) 1D bucket signals across different spectral channels.}
\label{f1}
\end{figure}

Our single-pixel HSI experimental apparatus is depicted in Fig.~\ref{f1}, which is similar to the typical SPI scheme, but with the use of a spectrometer instead of a single-pixel detector. An optical beam from a near-infrared lamp was projected onto an object $O(u,v,\lambda)$, the reflected beam from which was imaging onto a digital micro-mirror device through an imaging lens. We employed the Hadamard matrix $H_m(u,v)$ as the measurement matrix to be loaded onto the DMD (see Fig.~\ref{f1}(a)), where $m=1,2,\ldots,M$, $M$ denoted the number of modulation pattens (i.e., the number of measurements). After modulation, one reflection beam of the DMD was converged onto a fiber spectrometer and dispersed into several spectral channels. The modulated signals were transformed into one-dimensional (1D) bucket signals of the same length (see Fig.~\ref{f1}(b)) \cite{20edgar2019principles}, denoted as $I_m=H_m(u,v)O(u,v,\lambda)$. To mitigate the impact of environmental noise, we used a differential GI (DGI) algorithm for image reconstruction, utilizing the following formula \cite{DGIferri2010differential}
\begin{equation}
O_{DGI}=DGI(H,I)=\langle H_m I_m\rangle-\frac{\langle H_m\rangle}{\langle S_m\rangle}\langle S_m I_m\rangle,\label{eq1}
\end{equation}
where $S_m=\sum_{(u,v)}H_m(u,v)$ denoted the pixel summation performed on the Hadamard pattern $H_m$, $O_{DGI}$ represented the recovered image of $N\times N$ pixels. Here, the SR was defined as $SR=m/N^2$.

\subsection{\label{sec:level2}Image reconstruction}
After data collection, we would use a physics-driven GAN framework to reconstruct the images.

In the SPI field, data-driven DL algorithms have proven effective in mitigating imaging challenges arising from ultra-low SRs, which is a common issue for conventional CS methods. Regrettably, obtaining sufficient training data is still a challenge in many tasks, and the limited model generalization remains a critical issue in real-world imaging \cite{fanhuaalzubaidi2023survey,fanhuabrigato2020close}. The introduction of physics-driven CNN methods provide new solutions for the model generalization issue, relying solely on the use of mean squared error (MSE) objective function.

Inspired by the above physics-driven mechanism, we proposed a physics-driven GAN approach for single-pixel infrared HSI reconstruction, where the physical process was integrated with the generator. The method optimized the parameters by using both the MSE objective function and the adversarial objective function, thereby enabling the recovery of high-quality images without the need for dataset training. This physics-driven GAN based image reconstruction method could be formulated as a min-max optimization problem (see Refs. \cite{zhu2017unpaired,45karim2021spi})
\begin{equation}
\begin{aligned}
\min_{G_\theta}\max_{D_\theta}V\left(D_\theta,G_\theta\right)=&E_{I_m\sim p\left(I_m\right)}\left[\log D_\theta\left(I_m\right)\right]\\
&+E_{z\sim p(z)}\left[\log\left(1-D_\theta\left(H_m G_\theta(z)\right)\right)\right].
\end{aligned}
\label{eq2}
\end{equation}
Here, $E_{r\sim p(r)}$ represented the mathematical expectation of $r$ ($r=I_m,z$), $z$ stood for the speckle map as the random input, $V(\ast)$ was the value function of the GAN, $G_\theta$ indicated the generator model for generating the estimated 1D bucket signals $H_m G_\theta(z)$, $D_\theta$ denoted the discriminator model for distinguishing between the estimated 1D bucket signal $H_m G_\theta(z)$ and the actual 1D bucket signal $I_m$. The discriminator needed to maximize its discriminative ability between $H_m G_\theta(z)$ (real) and $I_m$ (fake). The primary role of the GAN used here was to adjust the parameters of $G_\theta$ to make the estimated signal $\tilde{I}_m=H_m G_\theta(z)$ much closer to the real measured signal $I_m$. When $\tilde{I}_m$ and $I_m$ were very close to each other, $G_\theta(z)$ at this point is the final output reconstructed image.

As mentioned above, we would optimize the parameters of $G_\theta$ through the objective function, which consisted of parts from both the MSE objective function and the adversarial objective function. The MSE objective function was defined as follows \cite{45karim2021spi,creswell2018generative},
\begin{equation}
I_{\textrm{MSE}}=\left\|H_m G_\theta(z)-I_m\right\|^2.\label{eq3}
\end{equation}
\begin{figure}[htbp]
\centering
\includegraphics[width=1\linewidth]{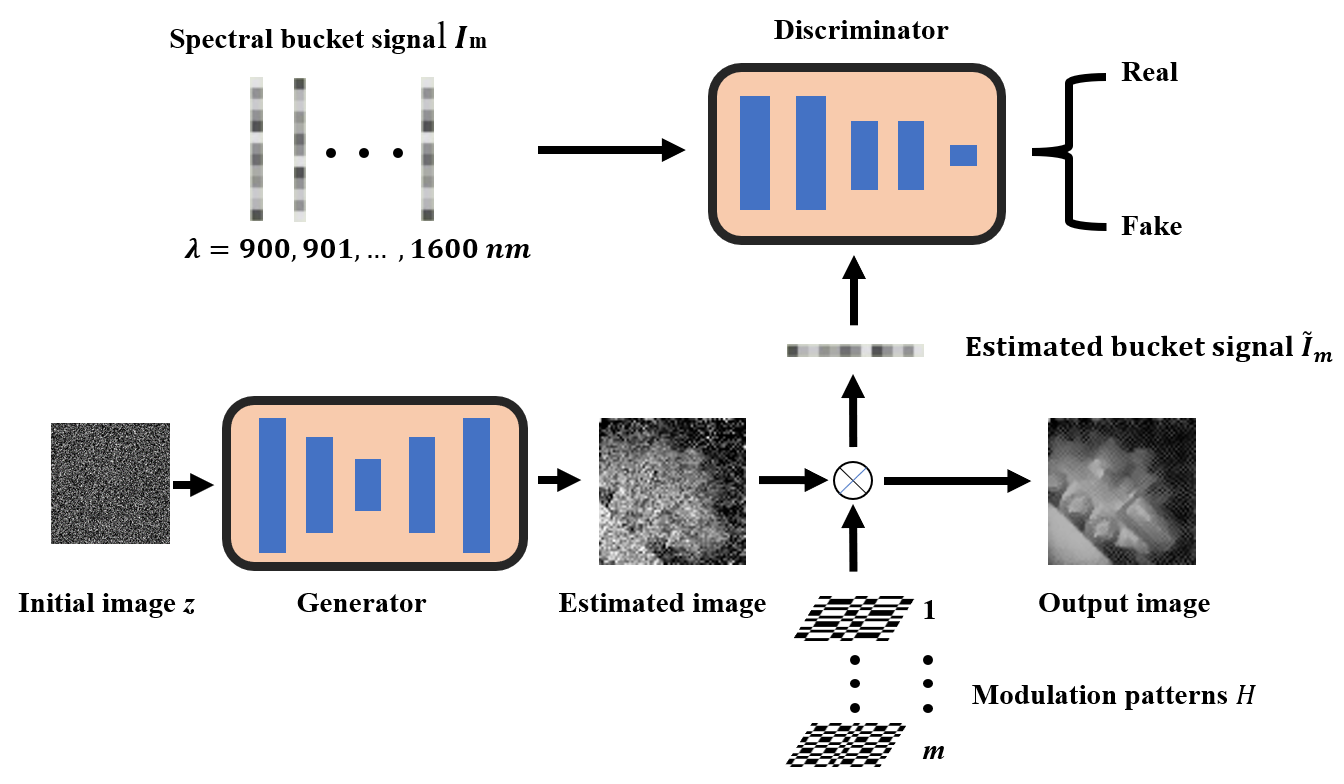}
\caption{The image reconstruction process by using the proposed physics-guided GAN method.}
\label{f2}
\end{figure}
In principle, there are infinitely many $G_\theta(z)$ that can satisfy the objective function. Hence, it is necessary to add prior information about the image to select a feasible solution from all options. However, existing work empirically demonstrated that an appropriately designed neural network with randomly initialized weights possesses a prior biased towards natural images \cite{52wang2020phase,53lin2022compressed,54gelvez2023mixture,55bostan2020deep}. Therefore, we also employed the randomly initialized prior in our generator network.

Next, we would use the objective function $l_{\textit{adv}}$ to optimize the network parameters. The purpose of the adversarial constraint was to make the output much closer to the real data. Since this objective function was related to the discriminator network, we minimized the negative log-likelihood of the discriminator, thereby increasing the probability that the generator's output values $H_m G_\theta(z)$ close to $I_m$. The adversarial objective function $l_{\textit{adv}}$ could be written as \cite{creswell2018generative}
\begin{equation}
l_{\textrm{advend}}=-\log D_\theta\left(H_m G_\theta(z)\right),
\label{eq4}
\end{equation}
where $G_\theta(z)$ was the output of the generator, and $D_\theta\left(H_m G_\theta(z)\right)$ signified the probability that the estimated 1D bucket signal $H_m G_\theta(z)$ and the actual 1D bucket signal $I_m$ were very close to each other. The final objective function was \begin{equation}
l=l_{\textit{MSE}}+l_{\textit{adv}}.
\label{eq5}
\end{equation}
The generator network parameters could be optimized through the following formula
\begin{equation}
G_\theta^\ast=\arg \min\left(\left\|H_m G_\theta(z)-I_m\right\|^2-\log D_\theta\left(H_m G_\theta(z)\right)\right),\label{eq6}
\end{equation}
where $\theta$ in $G_\theta^\ast$ represented the optimal parameters for the generator. By minimizing the objective function in Eq.~(\ref{eq6}), the optimal parameters for the generator were obtained, making $H_m G_\theta^\ast(z)$ much close to $I_m$. And at this point, the reconstruction $G_\theta^\ast(z)$ was at its best.

It's worth noting that SPI-based GAN \cite{45karim2021spi} also employs GAN but reconstructs images through the following formula
\begin{equation}
G_\theta^\ast=\arg\min\left(\frac{1}{WH}\sum_{i=1}^W \sum_{j=1}^H (O_s-G_\theta(z)_{i,j})^2-\log D_\theta(G_\theta(z)_{i,j})\right),\label{eq7}
\end{equation}
where $O_k$ is the ground truth image in the dataset. The SPI-based GAN learns the mapping from low-quality images to high-quality images in the dataset $s=\{(O_s^{\textit{DGI}},O_s)$, $s=1,2,\ldots,S$, to optimize the parameters of the generator $G_\theta$. Conversely, our proposed method updates the parameters in the neural network through the objective function constrained by $H_m G_\theta(z)$ and $I_m$, which can be seen as the combination between the GI physical model $H_m$ and $G_\theta$. In contrast, by using our approach, the reconstructed image $O_{\textit{Ours}}^\ast=G_\theta(O_{\textit{DGI}})$ was obtained without any dataset training, where $O_{\textit{Ours}}^\ast$ denoted the reconstructed image. It is also noteworthy that the input of our proposed neural network can be a rough image $G_\theta(O_{\textit{DGI}})$ restored by any traditional GI algorithm, or even a random noise image $G_\theta(z)$. Hereinafter, we used $G_\theta(z)$ for convenience.

\begin{figure}[htbp]
\centering
\includegraphics[width=1\linewidth]{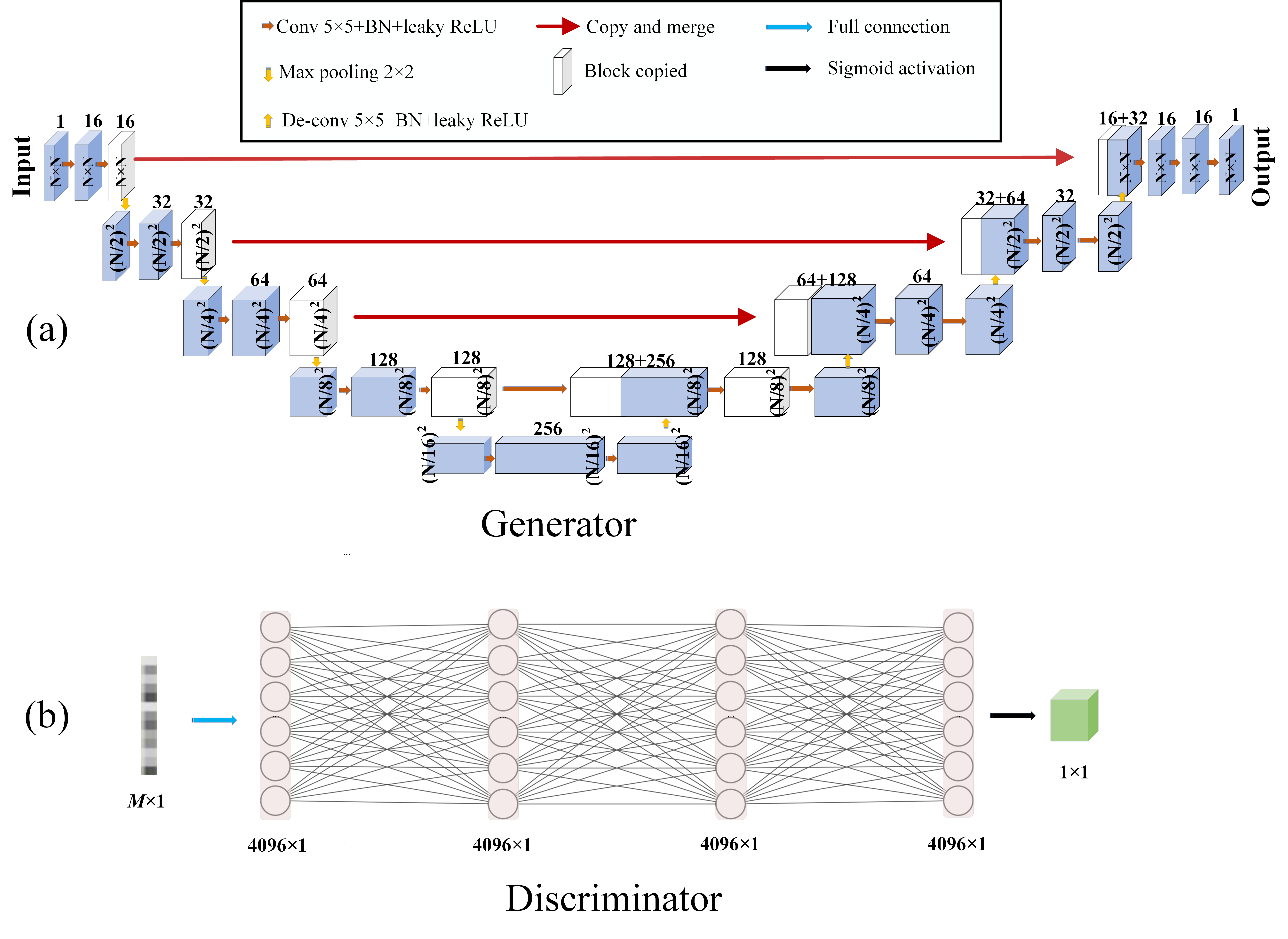}
\caption{The network structure of the proposed method, consisting of (a) generator network and (b) discriminator network.}
\label{f3}
\end{figure}

The principle process of the proposed method is described as follows (see Fig.~\ref{f2}). Step 1: the measured 1D bucket signal $I_m$ is input into the discriminator, which applies cross-entropy objective function for solving the maximum and minimum problem (Eq.~(\ref{eq2})). The parameters of the discriminator $D_{\theta}$ are updated by using the backpropagation algorithm, enhancing its ability to distinguish between real and fake data. Step 2: a random noise image $z$ is input into the untrained generator to produce the image $G_{\theta}(z)$, which is then converted into a 1D bucket signal $I$ by using the modulation patterns $H$, i.e., $I=HG_{\theta}(z)$. The objective function of the generator (Eq.~(\ref{eq6})) is computed and used to guide the update of the generator's parameters $G_{\theta}(z)$, also by employing the backpropagation algorithm. We need to execute Steps 1 and 2 in turn. By this means, the iterative process of the generative adversarial network can be regarded as a dynamic game process, where the generator and discriminator improve their performance through mutual antagonism and cooperation. Alternate iterations until the generator and discriminator reach an ideal performance level or the number of iterations reaches the set value.

The network framework of the proposed method is shown in Fig.~\ref{f3}, where the generator consists of two processes: up-sampling and down-sampling. In down-sampling process, there are convolution blocks ($5\times5$ convolution (stride 1) $+$ batch normalization (BN) + leaky ReLU) and $2\times2$ max pooling operations (stride 2). The up-sampling involves deconvolution blocks ($5\times5$ deconvolution (stride 2) + BN + leaky ReLU). The discriminator comprises four fully connected layers, each of $4090\times1$, ending with a sigmoid function. The number $N$ in the generator network is set to 64. Both networks use a learning rate (LR) of 0.005 and employ leaky ReLU as the activation function. Here, an Adam optimizer is used to optimize the network parameters.

\section{\label{sec:level1}Results and Discussion}
In order to verify the effectiveness of our method, we conducted a comparison with the CS and GIDC algorithms through simulations and experiments. We employed the peak signal-to-noise ratio (PSNR) (a commonly used metric) and structural similarity (SSIM) index to quantitatively analysing reconstruction quality. They all compare the differences between the reconstructed image and ground truth image. The formula for calculating PSNR is
\begin{equation}
\textrm{PSNR}=10\cdot\log_{10}(\frac{\text{MAX}_K^2}{\text{MSE}}),\label{eq8}
\end{equation}
where $\text{MAX}_K$ is the maximum value of the image, and MSE is expressed by
\begin{equation}
\text{MSE}=\frac{1}{xy}\sum_{a=1}^{x}\sum_{b=1}^{y}[J(a,b)-K(a,b)]^2.\label{eq9}
\end{equation}
Here, $J(a,b)$ and $K(a,b)$ represent the ground truth image and the reconstructed image, respectively, while $a$ and $b$ denote the row and column coordinates of images. The SSIM is a criterion for evaluating the similarity between two images, and is commonly used in the field of image processing to assess image quality. It is based on the computation of three terms that compare the luminance, contrast and structure of the test and reference images. The formula for the SSIM can be written as
\begin{equation}
\text{SSIM}(J,K)=\frac{(2\mu_J\mu_K+c_1)(2\sigma_{JK}+c_2)}{(\mu_J^2+\mu_K^2+c_1)(\sigma_J^2+\sigma_K^2+c_2)},\label{eq10}
\end{equation}
where $J$ and $K$ are the two images to be compared, $\mu_J$ and $\mu_K$ are their average luminances, $\sigma_J$ and $\sigma_K$ are their standard deviations, $\sigma_{JK}$ is the covariance between both images, and $c_1$ and $c_2$ are small constants for avoiding a zero denominator. The value of the SSIM ranges from -1 to 1, where 1 indicates perfect similarity between two images.

\subsection{\label{sec:level2}Simulations}
\begin{figure}[htbp]
\centering
\includegraphics[width=1\linewidth]{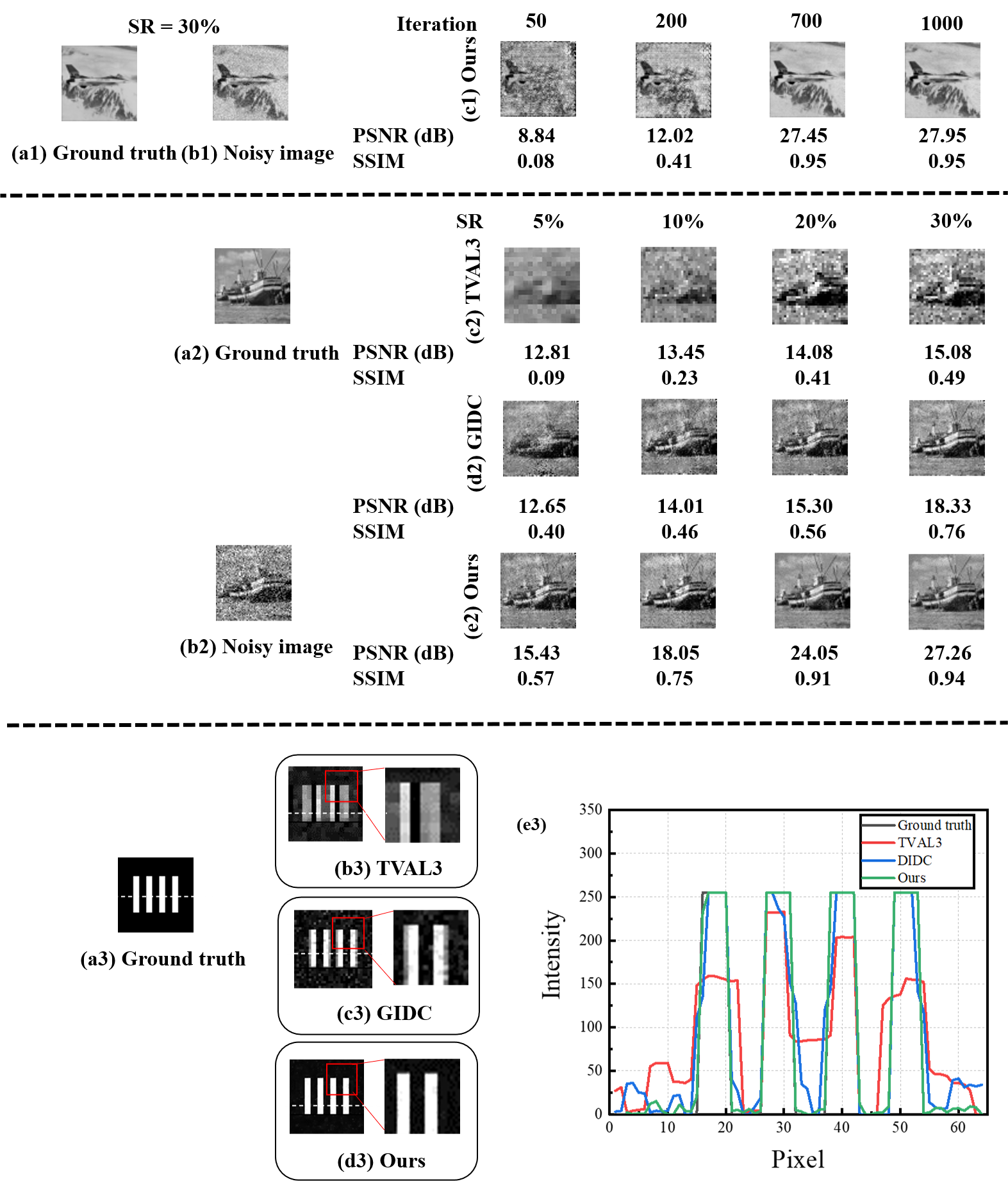}
\caption{Simulation test of the proposed method. (a1)--(a3) represented the ground truth images. (b1) and (b2) stood for the noisy images. (c1) showed the changes in the reconstruction quality of our method as the number of iterations increased. (c2)--(e2) represented the images recovered by the TVAL3 method, GIDC and our method as the sampling rate (SR) increased, respectively. (b3)--(d3) presented the image details retrieved by the TVAL3 method, GIDC and our method, respectively. (e3) gave the cross section plot of the restored images as shown in (b3)--(d3) on the same row to make comparison of details among these three methods.}
\label{f4}
\end{figure}

Since each wavelength channel in the single-pixel HSI is discrete, we first performed the numerical simulation for a single channel by using our physics-based GAN. Here, a gray-scale airplane image of $64\times64$ pixels (see Fig.~\ref{f4}(a1)) was used as the target to be imaged. To simulate the actual measurement noise, we added white Gaussian noise to the original image given in Fig.~\ref{f4}(a1), obtaining a noisy image as shown in Fig.~\ref{f4}(b1), which was considered as the image to be imaged onto the DMD. We applied $M=1229$ modulation patterns to generate the 1D bucket signal (which was of $1\times M$ in length), leading to the SR$=1229/4096\approx30$\%. The LR was set to 0.005. According to our proposed scheme, the physical process was simulated as follows. The Hadamard patterns were rearranged by following the Harr wavelet transform ranking method \cite{xiaoboli2019fast}. By this means, each modulated image was summed pixel by pixel to obtain the 1D bucket signal $I_m$ (Fig.~\ref{f1}(b)), which was fed into the discriminator, while the random noise map was treated as the input of the generator. By following the steps in the principle section and optimizing the parameters using the objective function (see Eqs.~(\ref{eq2}) and (\ref{eq7})), the optimal generator $G_{\theta}^\ast$ and reconstructed image $G_{\theta}(z)$ were obtained. We set the number of iterations to 50, 200, 700 and 1000, and gave their corresponding reconstructed images in Fig.~\ref{f4}(c1). It could be clearly seen that the image quality improves with the increase of the number of iterations, proven by PSNRs and SSIMs. When the number of iterations is less than 700, the quality improvement is obvious, and when the number of iterations is greater than 700, the quality improvement tends to be saturated, as the network tends to converge at this point.

To compare the performance of our network with other algorithms, another simulation was performed. Another boat image of $64\times64$ pixels (see Fig.~\ref{f4}(a2)) was used as the target and its corresponding noisy image was shown in Fig.~\ref{f4}(b2). The CS-based SPI (with well-known total variation augmented Lagrangian alternating direction (TVAL3) \cite{TVAL3li2010efficient} algorithm) and SPI-based CNN (with commonly used ``GI using deep neural network constraint" (GIDC) \cite{50GIDCwang2022,wang2023single} algorithm) were utilized for reconstruction under the same conditions. The SR was changing from 5\% to 30\%, and their corresponding results for the three methods were presented in Figs. \ref{f4}(c2) to \ref{f4}(e2). It could be clearly seen that the image quality all improved as the increase of SR. For each given SR, our method performed the best, followed by GIDC, and TAVL3 yielded the lowest quality. Based on the visual analysis of Figs.~\ref{f4}(c2)--\ref{f4}(e2),
it become apparent that the utilization of TVAL3 resulted in a notable decline in image quality at an SR of 10\%. Conversely, images reconstructed using the GIDC exhibited degradation when the sampling rate reached 5\%. In contrast, the images restored by using our proposed method remained perceptibly clear even when the SR decreased to 5\%.

In order to perform a quantitative analysis of the image quality, the PSNRs and SSIMs of images reconstructed by the above three methods at various SRs were also calculated and given in Figs.~\ref{f4}(c2)--\ref{f4}(e2). As the SR decreased, the PSNRs of the images recovered by all three methods showed an decreasing trend, which was consistent with the trend indicated in the images. At an SR of 5\%, the PSNR and SSIM of our method reached 15.45 dB and 0.57, respectively, while the PSNR and SSIM of TVAL3 and GIDC methods were only (12.65 dB, 0.40) and (12.81 dB, 0.09), respectively. It even surpassed the PSNRs of the images restored by TVAL3 at a 30\% SR and was almost equivalent to the images reconstructed by GIDC at a SR of 20\%. Based on the above analysis, it was evident that our method was proficient at reconstructing target images even at extremely low SRs, and it consistently yielded significantly better image quality compared with the other two methods. This exceptional performance underscored the promising potential of our proposed method for practical imaging scenarios.

\begin{figure}[htbp]
\centering
\includegraphics[width=1\linewidth]{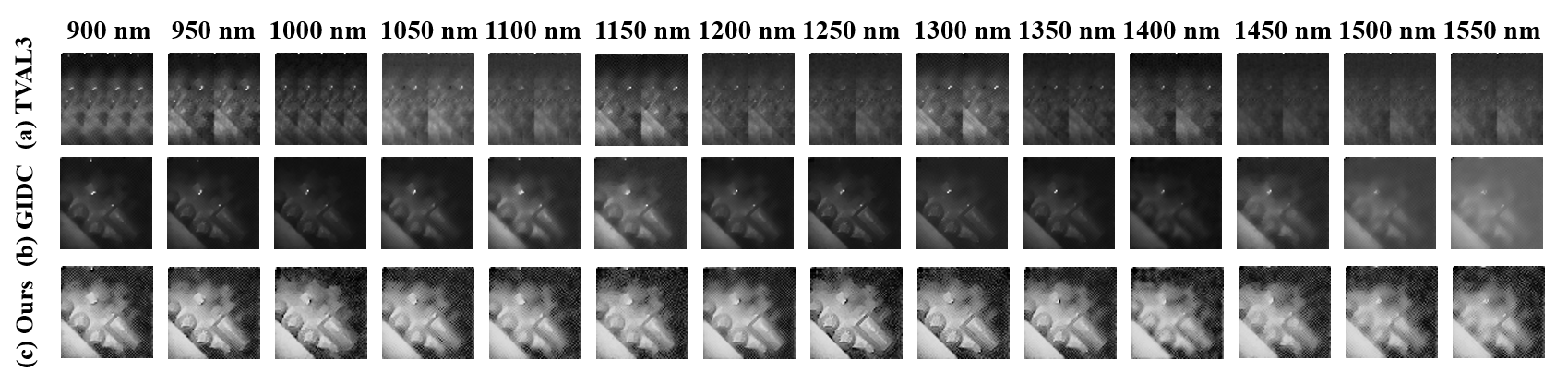}
\caption{Results of reconstructing the physical train model using different methods. (a)--(c) represented the reconstruction results obtained using the TVAL3, GIDC, and our method, respectively. From left to right, the wavelengths of the light ranged from 880~nm to 1600~nm with a step of 50~nm.}
\label{f5}
\end{figure}
\begin{figure}[htbp]
\centering
\includegraphics[width=1\linewidth]{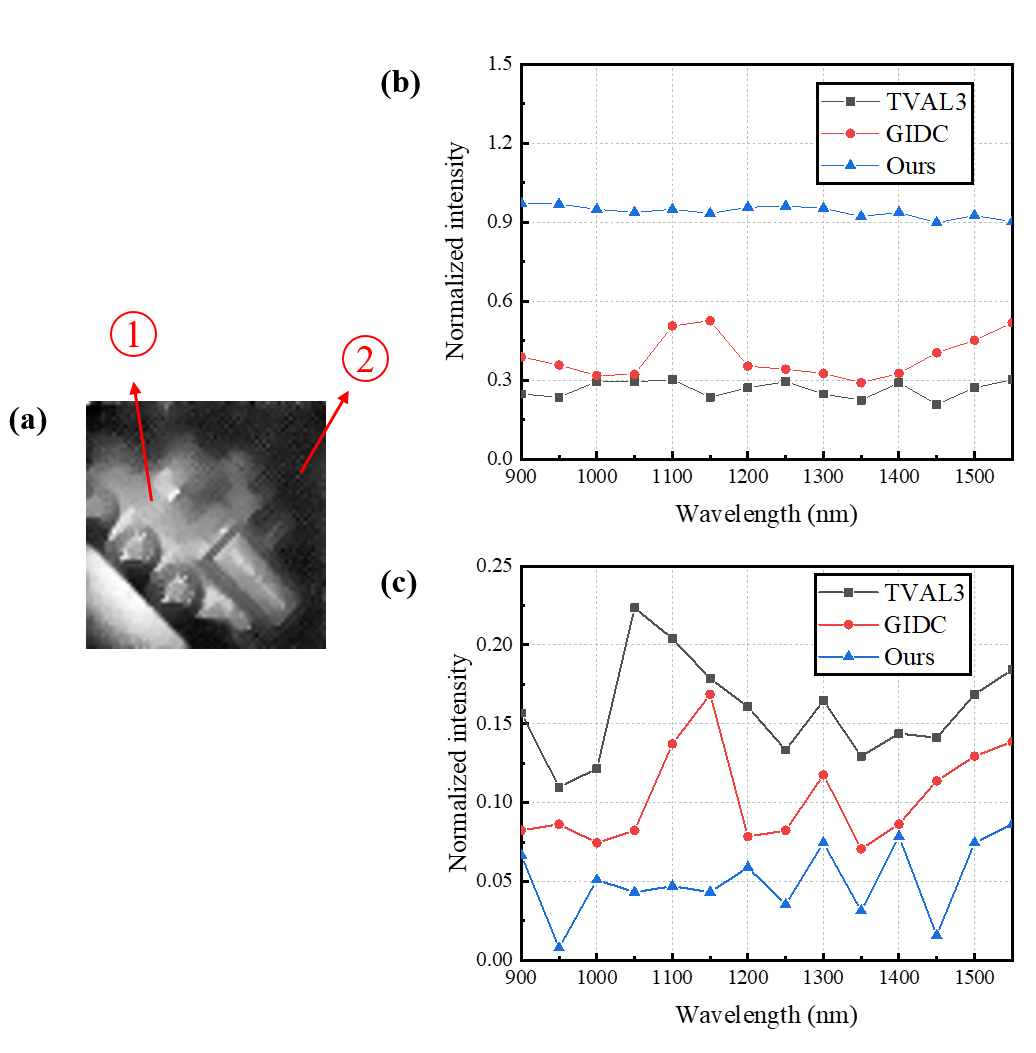}
\caption{Quantitative analysis results of single-pixel infrared HSI. (a) Two specific pixel locations were selected on the image, separately representing the reconstructed object part and the background part.
(b)--(c) represented the results obtained at pixel \textcircled{{\raisebox{-0.2ex}1}} and \textcircled{{\raisebox{-0.2ex}2}} by three methods at different wavelengths. The black squares, red circles, and blue triangles stood for the TVAL3 method, GIDC method, and the proposed method, respectively.}
\label{f6}
\end{figure}

Furthermore, an analysis of image details using three methods was provided through simulation to verify the enhancement of image quality. In this case, we used a binary $64\times64$ pixel object with four identical slits, each of 5 pixels in width and 30 pixels in length, and equally spaced 6 pixels apart, as shown in Fig.~\ref{f4}(a3), Other conditions were the same as the previous simulations. The simulated results at a 5\% SR were shown in Figs.~\ref{f4}(b3)--\ref{f4}(d3), respectively. Each retrieved image was accompanied by a zoomed-in image in the upper right corner. It could be clearly seen that the images reconstructed by TVAL3 exhibited prominent artifacts, as shown in Fig.~\ref{f4}(b3). However, in contrast, the image reconstructed by the GIDC method (see Fig.~\ref{f4}(c3)) displayed an enhancement in both quality and contrast, though it did not entirely eliminate the artifacts. Importantly, as illustrated in Fig.~\ref{f4}(d3), the image reconstructed by our proposed method effectively eliminated the artifacts, thereby demonstrating its superior imaging quality. Figure~\ref{f4}(e3) was a plot of the cross sections taken on the white dashed lines in Figs.~\ref{f4}(a3) and \ref{f4}(b3)--\ref{f4}(d3), where the black, red, blue and green lines represented the sections of the object and the images reconstructed through TVAL3, GIDC and our method, respectively. From these curves, it become strikingly clear that our proposed method excelled in reconstructing the object's details, showing the highest overlap with the original image's edge features. And the GIDC method followed with a notable performance, while the image reconstructed by the TVAL3 almost entirely lost the edge features and with poor quality. These results strongly highlighted the outstanding reconstruction abilities of our method.

\subsection{\label{sec:level2}Experiments}
In the previous subsection, we had demonstrated the outstanding performance of our physics-driven GAN method in SPI through a comparative analysis with two other image reconstruction techniques in numerical simulations. In this subsection, we continued to verify the performance of our physics-driven GAN in infrared single-pixel HSI experiments. The experimental setup was shown in Fig.~\ref{f1}, where we used a toy train model as the imaging target. An infrared beam emitted from a lamp with a wavelength range of 880~nm to 1600~nm illuminated the toy train, which was then imaged onto the DMD (ViALUXV-7001) through an imaging lens ($f=60$~cm) with a focal length of 60~mm. Then, after modulation, one of the beams reflected from the DMD passed through a collecting lens onto an infrared spectrometer (FUXIAN, NIR17+Px), and the recorded signal was dispersed into 512 spectra with a step of $\sim2$~nm. All the bucket signals, denoted as $I_m$, extracted from the spectrum were stored.

In our physics-driven GAN, a random noise image $z$ was used as the input for the generator, the result $G_\theta(z)$ of which was multiplied by modulation patterns $H_m$ to produce $\tilde{I}_m$. Simultaneously, the 1D bucket signals $I_m$ of 512 distinct wavelengths were sequentially fed into the discriminator network. The SR was set as 25\%. Upon the network's completion, a total of 512 near-infrared hyperspectral images were sequentially reconstructed. For comparative purposes, we utilized the TVAL3 and GIDC methods to reconstruct an equivalent set of 512 near-infrared hyperspectral images under identical conditions. To ensure a concise presentation, we deliberately selected only 14 images for each method, spaced at 50~nm intervals. These selected images are showcased in Fig.~\ref{f5} to underscore the key findings. The image reconstructed by TVAL3 were indistinguishable. On the other hand, the GIDC method successfully reconstructed the train model image, but it still struggled with image clarity within the wavelength range of 900~nm to 1000~nm and 1200~nm to 1400~nm. In contrast, our proposed method consistently produced reconstructed images with significantly enhanced contrast and clarity across all wavelength ranges. Consequently, when compared to the other two methods, our proposed approach demonstrated superior imaging performance at low SRs in the single-pixel infrared HSI scheme.

To further assess the performance of our proposed single-pixel infrared HSI method, we conducted a quantitative analysis. We obtained reconstructed images at various wavelengths by using three different methods. As shown in Fig.~\ref{f6}, once the light intensity had been normalized, Pixel \ding{172} corresponded to the light transmission portion of the reconstructed object image, with an intensity of 1; while Pixel \ding{173} corresponded to the non-light transmission portion of the background image, with an intensity of 0. To evaluate the reconstructed image's quality, we compared the light intensity differences at the specific pixels for each method with actual light intensity. A closer match between the pixel values of the reconstructed image and actual values indicated a more accurate imaging performance. By this means, it allowed us to quantitatively evaluate and compare different methods in terms of reconstructed image quality, providing a more precise assessment of our proposed method's performance in single-pixel infrared HSI applications.

\begin{figure}[htbp]
\centering
\includegraphics[width=1\linewidth]{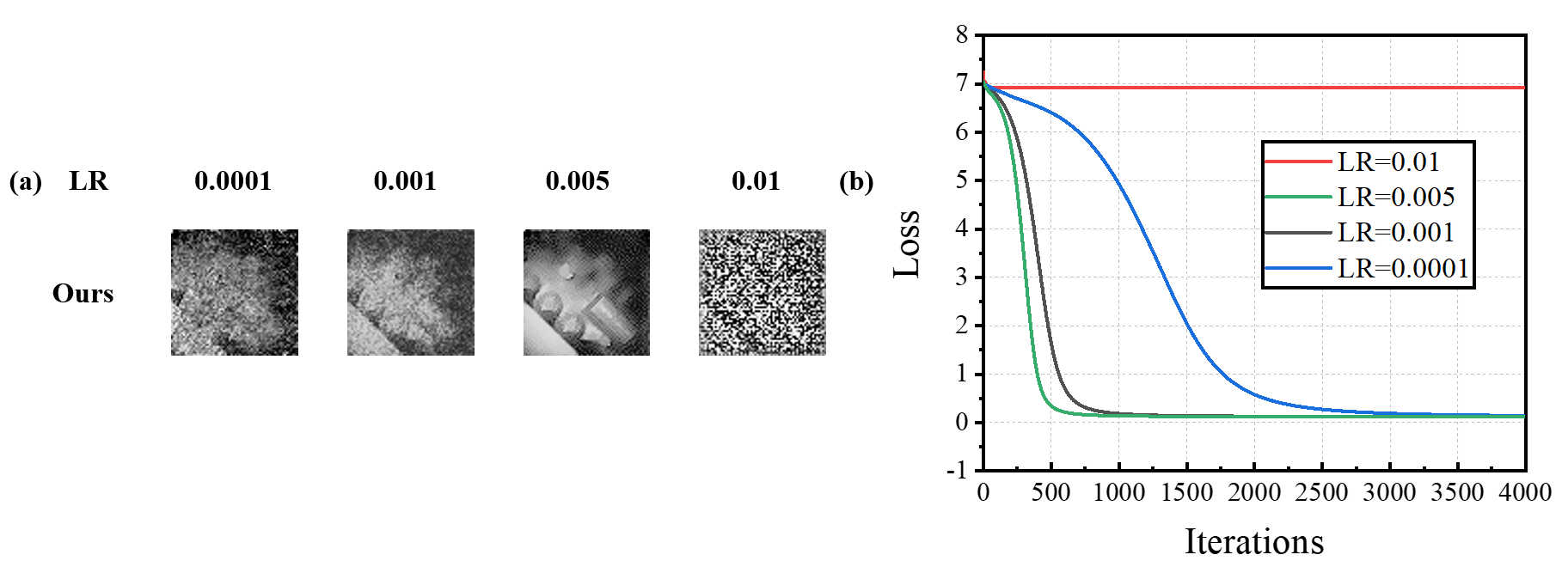}
\caption{Imaging performance of our method with 1000 iterations under different learning rates (LRs), (a) gave a comparison of reconstruction quality. (b) drew a plot of the loss function of the generator against the number of iterations with different LRs.
}
\label{f7}
\end{figure}
\begin{figure}[htbp]
\centering
\includegraphics[width=1\linewidth]{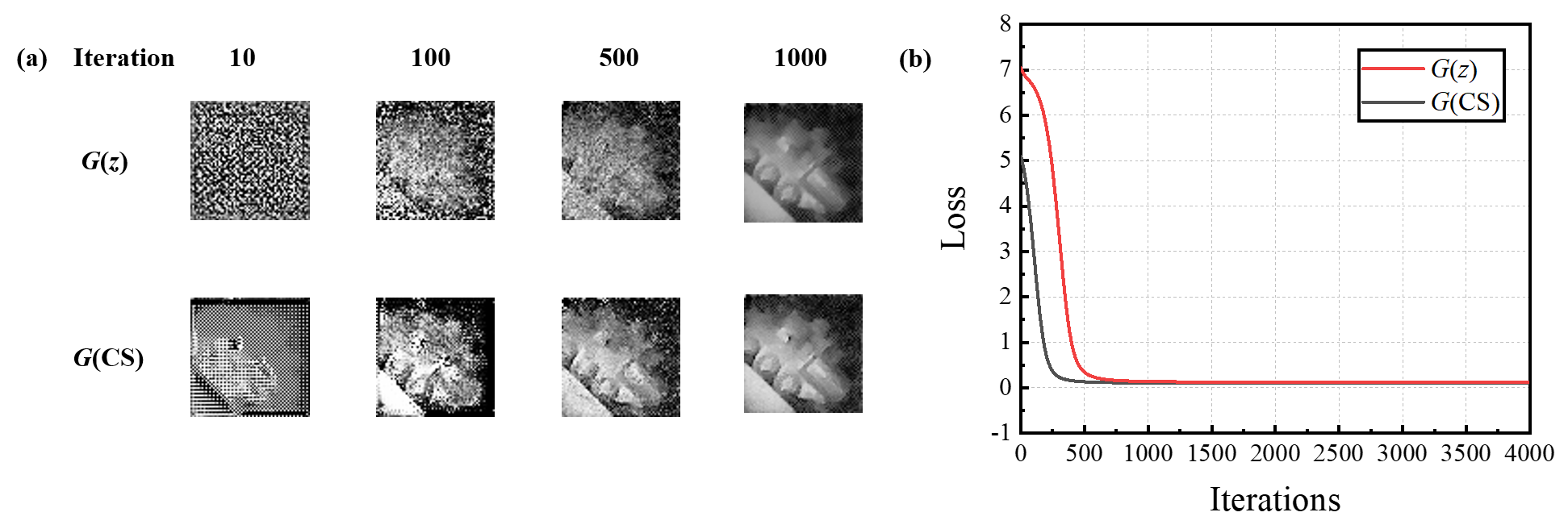}
\caption{Comparison of using different initial images. (a) gave comparison of applying the images reconstructed by the CS algorithm and the random noise images as the inputs of the generator, using different numbers of iterations. (b) drew curves of the generator's loss as a function of the number of iterations.
}
\label{f8}
\end{figure}

In the preceding paragraphs, we compared the experimental results of three different methods, demonstrating convincingly that our proposed physics-driven GAN could achieve better image quality under the same conditions. As it is known that the configuration of network parameters also plays a crucial role in the effectiveness of image recovery. Therefore, we further demonstrated this point by investigating the network parameters. Here, We set SR to 0.25\% and the number of network iterations as 1000. In our network, the LR is an important parameter. It could be observed from Fig.~\ref{f7}(a) that the images recovered with LRs of 0.0001 and 0.001 were relatively blurred, whereas the recovered image with LR of 0.005 exhibited a better image quality. However, when the LR increased to 0.01, the retrieved images become completely blurred again. This could be explained through Fig.~\ref{f7}(b). It could be observed that when the LR was 0.0001, which was too low, resulting in a slower convergence of the generator's loss function and requiring more iterations. At a LR of 0.001, the convergence speed increased, requiring fewer iterations. When the LR was 0.005, the convergence was the fastest, requiring the least number of iterations. However, when the LR reached 0.01, the loss function failed to converge, resulting in the network not functioning properly and being unable to clearly reconstruct images. Therefore, appropriate setting of the LR is crucial not only for image quality but also for reducing the number of iterations, accelerating network convergence, and improving imaging speed.

Next, we would further investigate the influence of the initial input image of the network \cite{50GIDCwang2022}. As mentioned in the second section of this paper, the input of the generator could be chosen as any image including a rough image restored by a traditional GI algorithm and even a random noise image $z$. Here, we selected the images reconstructed by the CS algorithm and random noise images as the input of the generator for comparisons, using different numbers of iterations, with a LR of 0.005 and a SR of 25\%. Here, $G(z)$ and $G(\textrm{CS})$ represented the results using random noise image and the CS recovered image as the input, respectively. The number of iterations was set to 10, 100, 500 and 1000, and the corresponding results were given in Fig.~\ref{f8}(a). It could be seen that $G(\textrm{CS})$ outperformed $G(z)$ at the number of iterations of 10, 100, and 500. The reason for this was that the corresponding loss function converged much faster with the CS-recovered image as input than with the random image as input. When the number of iterations reached 1000, they all saturated to the same loss value, as plotted in Fig.~\ref{f8}(b). The difference between these two initial inputs was due to the fact that in the early stages of iterations, the generator made fewer updates to the input image, while the restored image input already has a rough outline of the target image before being treated as the input. This also confirmed that the image input to the generator could be any image.

\section{\label{sec:level1}Conclusions}
We proposed a physics-driven GAN framework, which was successfully applied to single-pixel infrared HSI. Over 500 spectral images of the imaging target have been achieved at a 25\% SR. The fusion of physical processes and the GAN primarily exhibits two advantages. First, through 1D bucket signals, we are able to impose constraints on the MSE objective function and adversarial objective function of the GAN, thereby updating the network parameters and eliminating the dependency on datasets for training, which significantly enhances the generalization capability of the network. Second, the constraints of multi-objective functions contribute to a notable improvement in the quality of the reconstructed images. Our numerical simulations and single-pixel infrared HSI experiments have provided strong evidence for the excellent performance of the physics-driven GAN in achieving high-quality image reconstruction at low SRs. This forms a robust foundation for its application in areas such as infrared spectral detection. Although such a configuration might incur some time consumption, it enables a broader spectral detection range and lower costs.

\

\noindent\textbf{Funding.} National Key Research and Development Program of China (2018YFB0504302); Beijing Natural Science Foundation (4222016); National Defence Science and Technology Innovation Zone (23-TQ09-41-TS-01-011).

\

\noindent \textbf{Disclosures.} The authors declare no conflicts of interest.

\

\noindent \textbf{Data availability.} Data underlying the results presented in this paper are not publicly available at this time but may
be obtained from the authors upon reasonable request.

\bibliographystyle{unsrt}
\bibliography{ref}
\end{document}